\documentclass[aps,superscriptaddress,superbib,twocolumn,groupedaddress]{revtex4}

\bibpunct{}{}{,}{s}{}{}

\usepackage[french]{babel}
\usepackage[cyr]{aeguill}
\usepackage{amsmath,amssymb,amsfonts}
\usepackage{color}
\usepackage{graphicx}
\graphicspath{{fig/}}

\usepackage{dcolumn}
\usepackage{bm}
\usepackage{units}
\usepackage{tabularx}
\usepackage{version}

\renewcommand{\.}{\cdot}

\renewcommand{\d}{\partial}

\newcommand{\n}{\mathbf{n}}
\renewcommand{\r}{\mathbf{r}}

\renewcommand{\H}{\mathbf{H}}
\newcommand{\M}{\mathbf{M}}
\newcommand{\Oo}{\Omega}
\newcommand{\G}{\mathsf{G}}

\newcommand{\N}{$N$}
\renewcommand{\No}{$N$ }

\newcommand{\NlogN}{$N\log N$}
\newcommand{\NlogNo}{$N\log N$ }

\newcommand{\sqrtN}{$N^{1/2}$ }
\newcommand{\Ndutri}{$N^{2/3}$ }
\newcommand{\Ntetri}{$N^{4/3}$ }
\newcommand{\Ntridu}{$N^{3/2}$ }

\newcommand{\Nsqo}{$N^2$ }
\newcommand{\Nb}{$N_\d$ }
\newcommand{\Nbsq}{$N_\d^2$ }

\newcommand{\kt}[1]{\raisebox{-0.5ex}[0cm][0cm]{#1}}
\newcommand{\pn}[1]{\raisebox{+0.6ex}[0cm][0cm]{#1}}

\newcolumntype{Y}{>{\centering\arraybackslash$}X<{$}}
\newcommand{\mcol}[3]{\multicolumn{#1}{#2}{#3}}
\newcommand{\ti}[1]{\scriptstyle{#1}}
\newcommand{\tik}[1]{\ti{\text{\kt{#1}}}}
\newcommand{\tip}[1]{\ti{\text{\pn{#1}}}}

\begin{document}

\title{Fast computation of magnetostatic fields by Non-uniform Fast Fourier Transforms}

\author{Evaggelos Kritsikis}
\email{evaggelos.kritsikis@grenoble.cnrs.fr}
\affiliation{Spintec, CNRS/CEA/UJF/INP Grenoble, 38054 Grenoble cedex
9, France}
\author{Jean-Christophe Toussaint}
\affiliation{Institut N\'{e}el, CNRS et Universit\'{e} Joseph Fourier, BP 166, F-38042 Grenoble Cedex 9, France}
\affiliation{INP Grenoble, 38031 Grenoble cedex 1, France}
\author{Olivier Fruchart}
\affiliation{Institut N\'{e}el, CNRS et Universit\'{e} Joseph Fourier, BP 166, F-38042 Grenoble Cedex 9, France}

\date{\today}

\begin{abstract}

The bottleneck of micromagnetic simulations is the computation of the long-ranged magnetostatic
fields. This can be tackled on regular \N-node grids with Fast Fourier Transforms in time \NlogN,
whereas the geometrically more versatile finite element methods~(FEM) are bounded to \Ntetri in
the best case. We report the implementation of a Non-uniform Fast Fourier Transform algorithm
which brings a \NlogNo convergence to FEM, with no loss of accuracy in the results.

\end{abstract}


\maketitle

\vskip 0.5in

\vskip 0.5in


The power of computers steadily increases over the years while the size of devices used in
fundamental science or technology is shrinking. Today we have reached a cross-over where numerical
simulations are capable of describing in detail the physics of nanodevices, for which they thus
play a leading role in their understanding and designing. In numerical micromagnetics for spin
electronics, a bottleneck is the computation of the magnetostatic interactions which by nature
are long-ranged. These interactions can be expressed in terms of either magnetostatic field
$\H$ or scalar pseudo-potential $\phi$ such that $\H=-\nabla\phi$. The latter is convenient since
it boils the problem down to a single scalar unknown.
To deal with magnetostatic interactions, essentially two distinct approaches have been implemented,
depending on the type of mesh used:\\

\begin{enumerate}
\item for Finite Difference (ie. translation invariant) meshes, a Green approach with Fast Fourier
Transforms, called FD-FFT.
Computation time is moderate (\NlogNo with \No the number of nodes), but the curved
boundaries that occur often in experimental devices are not ideally described;\\
\item for Finite Element (much more general) meshes, a Finite Element Method coupled to a
Boundary Element Method, called FEM-BEM.
This can faithfully describe curved boundaries but computation time is higher,
at least \Ntridu in 2D (resp. \Ntetri in 3D).\\
\end{enumerate}

In this Letter we report the implementation of a new magnetostatic code which combines
the advantages of both cited approaches: it uses a FEM mesh thus describing curved
boundaries as well as FEM-BEM, albeit with computation time \NlogN. It is based on a algorithm
reported recently for computing non-periodic Fast-Fourier Transforms (NFFT) \cite{Pot01}.
Our code, called FEM-NFFT, proves to be significantly faster than FEM-BEM
with no loss of precision. As a first step the implementation was done for a 2D geometry
(which pertains to 3D systems with one direction of translational invariance, i.e. cylinder-like)
for the proof of concept. The gain is expected to be even greater in more realistic 3D
calculations. This demonstrates the potential of FEM-NFFT for micromagnetism and thus spin-electronic devices.

Let us recall the principle and features of FD-FFT and FEM-BEM before presenting our approach
and results. We consider a system $\Oo$ with boundary $\d\Oo$, displaying a known magnetization
distribution $\M(\r)$.\\

Finite Difference micromagnetic codes use a translation invariant grid.
On such a grid, FFTs can be used to compute convolutions in time \NlogN.
This motivates a Green approach for magnetostatics: $\phi$ is calculated as a
convolution of the Green function $\G = -(1/2\pi) \log\r$ in 2D [resp. $\G=1/(4\pi\r)$ in 3D]
with the magnetic charges, volumic $\rho=-\nabla\.\M$ and surfacic $\sigma=\M\.\n$:
\begin{equation}
  \label{green}
\phi(\r) = \int_\Oo \rho(\r')\.\G(\r-\r') \: \mathrm{d}\r' + \int_{\d\Oo} \sigma(\r')\.\G(\r-\r') \: \mathrm{d}\r'
\end{equation}

The \NlogNo speed explains the wide and lasting use of FD in micromagnetic simulations \cite{Kev98}.
However, most devices have curved boundaries, either by design or as a result of experimental
imperfections. Simulating these cases with FD requires the use of saw-tooth boundaries
to describe the magnetic material. This geometrical approximation may induce inadequate
descriptions \cite{Gar03}.\\

Finite Element micromagnetic codes use, on the contrary, complex-shaped meshes with triangles
(resp. tetrahedrons) as 2D (resp. 3D) unit cells. They consequently suffer much less from the
above-mentioned limitations. However, without translational invariance of the mesh, FFTs are not
available. Bearing in mind that direct summation of Eq.\eqref{green} on a FEM mesh,
called FEM-direct, would cost \Nsqo time, one understands why the Green approach is thought
incompatible with FEM codes.

To deal with magnetostatics, FEM codes thus go back to the Poisson equation $-\Delta\phi = \rho$,
with the usual regularity condition that the field should decay at infinity.
Applying FEM, the equations are translated into a linear system, which is solved by standard
iterative methods \cite{Bra01} in time \Ntridu in 2D (resp. \Ntetri in 3D)
Not only is this asymptotically slower than the \NlogNo time required for FD-FFT, but \No
here takes higher values, because the mesh must extend well beyond $\Oo$ in order to tackle
the regularity condition at infinity. This induces an additional slowdown, and also creates
finite-size artifacts.

To avoid meshing outside $\Oo$, the main approach consists in coupling FEM with a Boundary Element
Method, resulting in the so-called FEM-BEM \cite{FRE90}. The asymptotic complexity of the Poisson solver
is unchanged but the BEM step introduces another time limitation \Nbsq where \Nb is the number
of boundary nodes.
In the most favorable case consisting of compact systems \Nb $\approx$ \sqrtN in 2D (resp.
\Nb $\approx$ \Ndutri in 3D). However for flat geometries, of particular relevance to applications,
\Nb $\approx$ \N, in which case the time limitation for FEM-BEM may be pretty severe.\\

Our innovation is to revert, within the FEM framework, to a Green approach.
The convolution~\eqref{green} is discretized in a way typical for FEM, and computed using
a fairly recent mathematical method called NFFT (Non-uniform Fast Fourier Transform).
NFFTs allows one to compute discrete convolutions in time \NlogNo without the equispaced
data requirement of FFTs.

More in details, we seek $\phi$ at the nodes $(\r_i),\: \ti{i=1\ldots N}$ of the mesh.
A linear interpolation inside each element, of known magnetization values $\M(\r_i)$, is used
to evaluate charges $\rho$ and $\sigma$ at points $\r_j,\: \ti{j=1\ldots M}$ defined as the quadrature
points for the integrals in~\eqref{green}\cite{nfft-note1}. 

Consequently, \eqref{green} is rewritten the following way:
\begin{multline}
\label{greend}
\phi(\r_i) = \sum_{j=1}^M  \rho_j \: \G(\r_i-\r_j) \; \omega_j \: \det J(r_j)\\
  +  \sum_{j=1}^M \sigma_j \: \G(\r_i-\r_j) \; \omega_j \: \det J(r_j)
\end{multline}

\noindent where

\begin{itemize}
\item $\rho_j = \rho(r_j)$ if $r_j$ is in the interior of $\Oo$, $0$ otherwise,\\
\item $\sigma_j = \sigma(r_j)$ if $r_j$ is on $\d\Oo$, $0$ otherwise,\\
\item $\omega_j$ is the weight of $r_j$ in the quadrature scheme,\\
\item $J(r_j)$ is the Jacobian of the affine transformation mapping the unit element on that containing $r_j$.\\
\end{itemize}

We then use NFFTs to compute~\eqref{greend}.
Although the seminal paper \cite{Dut93} dates back to 1993,
the NFFT method remains little-known even in the mathematical community. A presentation
of the method can be found Ref.\cite{Gre04}. Here we sketch the basic strategy and show that
computation time does not exceed \NlogN, ie. that of the classical FFT.

Let us look at our non equispaced data as a sum of Dirac functions. The goal is to find
the spectrum of this data function. The idea is to \emph{convey initial information over a
regular grid}, so that an FFT can be used. We therefore choose a regular grid X and let the data
diffuse to the X nodes through convolution with a Gaussian function
(or more generally a smooth localized function).
The Gaussian is localized in space, so we can consider that each piece of data diffuses only to
a fixed number of the nearest X nodes. Computation time of this diffusion step is thus
proportional to \N.

The question arises how to choose the period of the X grid. In our case, the answer depends on
how smooth the Green function $\G$ is. A necessary preliminary step before executing the NFFT
is therefore to smooth $\G$ around the origin; the price to pay is an afterwards correction
in the smoothing zone. It can be shown that a grid of size $p^2 N$ in 2D (resp. $p^3 N$ in 3D)
is convenient, where $p$ is the degree of smoothness chosen for $\G$.

We then perform, according to the initial idea, a FFT on the X grid, in time proportional
to \NlogN. Based on the convolution theorem, what we get is the Fourier coefficients of the data
function \emph{multiplied} by those of the Gaussian. Therefore, we finally divide these numbers
by the Fourier coefficients of the Gaussian to get the desired spectrum. The number of divisions
is proportional to \N. As a whole, the NFFT is expected to behave asymptotically like \NlogN,
as all extra steps behave like \N.\\


We implemented FEM-NFFT to 2D test cases where an analytical solution $\phi_{\mathrm a}$ is available,
so that errors can be readily estimated.
Interpolation and quadrature routines are written in C++ and the NFFT package used \cite{Kun02} is
in C99. For $\G$ we have chosen a smoothness degree of 2. On each test case, we provide
computation times and error estimates for FEM-direct, the classic FEM-BEM and our NFFT-based method.
The computed error is the normalized root mean square
$(M_s\:L)^{-1} (\sum_{i=1}^N |\phi(\r_i)-\phi_{\mathrm a}(\r_i)|^2 / N)^{1/2}$,
where $M_s$ is the saturation magnetization and $L$ the system diameter set at unity, .
Computations are done on an Intel P4 2GHz with 1GB RAM running Fedora 5.

The first test case is a disk uniformly magnetized along the x-axis (a cylinder in 3D space).
The analytical solution
is $\phi(x)=M_s \: x /2$. The second case is the so-called magic cylinder, a circular annulus
of radii $R_1, R_2$ where the angle between magnetization and the x-axis equals twice the polar
angle. The name stems from the uniform magnetic field thus induced in the inner region. The analytical
solution inside the annulus is $\phi(r,\theta)= M_s \: r \cos\theta\log(r/R_2)$ in polar coordinates.

Tables 1 and 2 display the numerical results for the two cases, respectively.
It can be readily seen that FEM-NFFT provides results very similar to FEM-direct. The error
induced by the NFFTs is thus negligible. Compared to FEM-BEM, errors are comparable for
non-uniformly magnetized systems (see Table 2) whereas for uniform distributions,
Green approaches, to which FEM-NFFT belong, are more accurate by one order of magnitude
(see Table 1). This is because they can treate apart volumic and surfacic charge contributions.
Concerning computational time, FEM-direct is as expected the quickest, gaining a factor around 5 over FEM-BEM for the finer meshes.
Nodes required in 3D cases of interest commonly count up to $10^5$, around which number the time advantage of FEM-NFFT over classical
methods such as FEM-BEM is expected to reach one order of magnitude.

\begin{figure}[h]
\centering
\includegraphics[width=\linewidth]{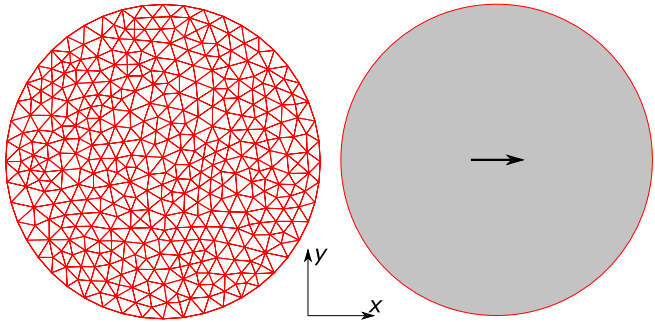}
\caption{Test case 1: the uniformly magnetized disk. Mesh used for $N=400$ (left) and magnetization distribution (right).}
\end{figure}

\begin{table}[h]
\caption{Test case 1: error (in ppm) and computation time (in seconds) of FEM-direct, FEM-BEM
and FEM-NFFT for different mesh sizes.}

\begin{tabularx}{\linewidth}{| Y !{\vrule width 2pt} Y|Y|Y !{\vrule width 2pt} Y|Y|Y|}
\hline
	&  \mcol{3}{c !{\vrule width 2pt}}{\textbf{error}}
	  &  \mcol{3}{c|}{\textbf{time}}							\\
\cline{2-7}
N	& \tik{FEM-}	& \tik{FEM-}	& \tik{FEM-}	& \tik{FEM-}	& \tik{FEM-}	& \tik{FEM}	\\
	& \tip{direct}	& \tip{BEM}	& \tip{NFFT}	& \tip{direct}	& \tip{BEM}	& \tip{NFFT}	\\
\hline
  400	&	93.6	&	162	&	90.3	&	0.401	&	0.070	&	0.089	\\
 1572	&	17.3	&	52.8	&	16.8	&	7.71	&	0.390	&	0.385	\\
 9489	&	2.16	&	25.8	&	2.09	&	233	&	4.75	&	2.18	\\
37938	&	 --	&	23.2	&	0.650	&   3720^\ast	&	40.1	&	9.81	\\	
\hline
\end{tabularx}
\flushleft{$\ti{^\ast:\text{ estimated}}.$}
\end{table}

\begin{figure}[h]
\centering
\includegraphics[width=\linewidth]{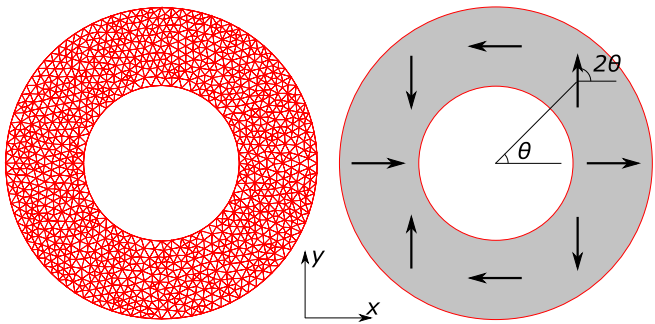}
\caption{Test case 2: the magic cylinder. Mesh used for $N=1192$ (left) and magnetization distribution (right).}
\end{figure}

\begin{table}[h]
\caption{Test case 2: error (in ppm) and computation time (in seconds) of FEM-direct, FEM-BEM
and FEM-NFFT for different mesh sizes.}

\begin{tabularx}{\linewidth}{| Y !{\vrule width 2pt} Y|Y|Y !{\vrule width 2pt} Y|Y|Y|}
\hline
	&  \mcol{3}{c !{\vrule width 2pt}}{\textbf{error}}
	  &  \mcol{3}{c|}{\textbf{time}}							\\
\cline{2-7}
N	& \tik{FEM-}	& \tik{FEM-}	& \tik{FEM-}	& \tik{FEM-}	& \tik{FEM-}	& \tik{FEM}	\\
	& \tip{direct}	& \tip{BEM}	& \tip{NFFT}	& \tip{direct}	& \tip{BEM}	& \tip{NFFT}	\\
\hline
 1192	&	174	&	172	&	174	&	3.58	&	0.290	&	0.331	\\
 2091	&	95.8	&	94.2	&	95.8	&	14.0	&	0.640	&	0.498	\\
 8214	&	24.7	&	32.1	&	24.8	&	211	&	4.59	&	2.02	\\
22683	&	 --	&	21.5	&	9.05	&   1600^\ast	&	34.0	&	6.62	\\	
\hline
\end{tabularx}
\flushleft{$\ti{^\ast:\text{ estimated}}.$}
\end{table}


To conclude, we have successfully implemented a Non-uniform Fast Fourier Transform~(NFFT) algorithm
to compute magnetostatic fields for micromagnetic simulations based on Finite Element methods~(FEM).
The new approach, called FEM-NFFT, combines the advantages previously found separately in Finite
Difference methods (computation time scaling like \NlogN) and FEM (faithful
description of curved boundaries). Thus FEM-NFFT promises a leap in the attractiveness of
micromagnetic simulations of spin electronic devices.


\begin{thebibliography}{9}
\bibitem[{nff()}]{nfft-note1}
\bibinfo{note}{In singular zones where the quadrature scheme becomes not
  precise enough, it is possible to combine it with semi-analytical terms, the
  details of which will come in a forthcoming paper}.
\end{thebibliography}

\end{document}